# Earth: Atmospheric Evolution of a Habitable Planet

**Stephanie L. Olson**[1,2]*****, Edward W. Schwieterman[1,2],
Christopher T. Reinhard[1,3], Timothy W. Lyons[1,2]

[1]NASA Astrobiology Institute Alternative Earth's Team
[2]Department of Earth Sciences, University of California, Riverside
[3]School of Earth and Atmospheric Science, Georgia Institute of Technology

*Correspondence: solso002@ucr.edu

**Table of Contents**

**Abstract** Our present-day atmosphere is often used as an analog for potentially habitable exoplanets, but Earth's atmosphere has changed dramatically throughout its 4.5-billion-year history. For example, molecular oxygen is abundant in the atmosphere today but was absent on the early Earth. Meanwhile, the physical and chemical evolution of Earth's atmosphere has also resulted in major swings in surface temperature, at times resulting in extreme glaciation or warm greenhouse climates. Despite this dynamic and occasionally dramatic history, the Earth has been persistently habitable—and, in fact, inhabited—for roughly 4 billion years. Understanding Earth's momentous changes and its enduring habitability is essential as a guide to the diversity of habitable planetary environments that may exist beyond our solar system and for ultimately recognizing spectroscopic fingerprints of life elsewhere in the Universe.

Here, we review long-term trends in the composition of Earth's atmosphere as it relates to both planetary habitability and inhabitation. We focus on gases that may serve as habitability markers ($CO_2$, $N_2$) or biosignatures ($CH_4$, $O_2$), especially as related to the redox evolution of the atmosphere and the coupled evolution of Earth's climate system. We emphasize that in the search for Earth-like planets we must be mindful that the example provided by the modern atmosphere merely represents a single snapshot of Earth's long-term evolution. In exploring the many former states of our own planet, we emphasize Earth's atmospheric evolution during the Archean, Proterozoic, and Phanerozoic eons, but we conclude with a brief discussion of potential atmospheric trajectories into the distant future, many millions to billions of years from now. All of these 'Alternative Earth' scenarios provide insight to the potential diversity of Earth-like, habitable, and inhabited worlds.

## 1. Introduction

Earth's atmosphere is dominantly $N_2$ (78%) and $O_2$ (21%) by volume today. This abundant $N_2$ provides the majority of Earth's surface pressure, which is critical for the stability of liquid water, while N is an essential nutrient for all life on Earth. High levels of $O_2$ support the metabolic demands of complex animal life as well as the production of significant ozone ($O_3$) in the stratosphere, which protects life on land from DNA-damaging UV radiation (*e.g.*, Catling et al 2005). Meanwhile, trace levels of $CO_2$, $CH_4$, and $H_2O$ warm the planet, resulting in a global average surface temperature of ~15°C—a full 33°C warmer than the planet would be without warming by greenhouse gases (Kump et al 2010).

However, none of these features of the Earth's atmosphere has been static throughout its history. The atmosphere has changed dramatically through time, both compositionally and physically. Whereas the present day Earth is strongly oxidizing, the early Earth was reducing and lacked atmospheric $O_2$ (Lyons et al 2014) and, consequently, a protective $O_3$ layer to prevent harmful UV irradiation of its surface. Earth has also experienced extreme temperature swings and long-lived, low-latitude glaciation as the result of major changes in the atmospheric abundance of greenhouse gases (Kasting 2005). Even Earth's predictably blue skies may have been a different color for extended intervals of Earth's early history due to the presence of hydrocarbon hazes (Arney et al 2016). In parallel with these atmospheric changes, Earth has experienced catastrophic impacts, violent volcanic eruptions, and mass extinctions (*e.g.,* Alvarez et al 1980). Yet the Earth has remained habitable, and inhabited, for at least the last 3.8 billion years (Nutman et al 2016).

Earth's present-day atmosphere ultimately arises from billions of years of co-evolution between the Sun, Earth's surface and interior, and Earth's biosphere. The current composition of our atmosphere does not represent a terminal atmospheric state, and as such it provides only a single snapshot along the long-term trajectory of the coupled evolution of the Earth system. Thus, in our search for life beyond our solar system, Earth provides many examples of Earth-like, habitable and inhabited planets—extending far beyond the template provided by our modern world.

Here, we explore the evolution of Earth's atmosphere through time and discuss the cause-and-effect relationships between these changes, the maintenance of habitability, and biological innovation. We focus our discussion on four biogeochemically important gases: $O_2$ (**Section 2**), $CO_2$ (**Section 3**), $CH_4$ (**Section 4**), and $N_2$ (**Section 5**). These gases have played important roles in the long-term operation of Earth's climate system, the maintenance of habitability, and Earth's capacity to support complex animal life. At the same time, the activities of Earth's biosphere have profoundly influenced the atmospheric abundance of each of these gases.

The first evidence for liquid water on Earth (*i.e.,* habitability) extends back to 4.4 billion years ago (Ga; Valley et al 2002). Because of the highly fragmented geologic record in the distant past, the timing of the origin of life remains unknown, but several authors have argued for early signs of life that extend back to 3.8 Ga (*e.g.*, Abramov and Mojzsis 2009; Mojzsis et al 1996)—and possibly as early as 4.1 Ga (Bell et al 2015). Consequently, the Hadean (>4.0 Ga) Earth may provide essential insight into how life can develop from non-life (see Sleep 2010), but there is a notable lack of reliable constraints on the Hadean atmosphere at present. For the balance of this review, we have restricted our discussion to the Archean, beginning at 4.0 Ga, and onward, thus focusing on the portions of Earth history for which we have geological constraints on surface chemistry as well as compelling evidence for inhabitation (*e.g.*, Nutman et al 2016).

## 2. Oxygen and biological innovation

Earth's early atmosphere was effectively devoid of $O_2$, with the exception of very low levels of abiogenic $O_2$ produced through photochemical reactions in the atmosphere (Kasting et al 1979). Although $O_2$ is an absolute requirement for complex animal life on Earth and likely elsewhere in the Universe (Catling et al 2005), $O_2$ is not a requirement for habitability. Instead, it may be our good fortune that the surface of the early Earth was anoxic. An oxidizing world would generally disfavor the prebiotic transformations among reduced compounds that culminated in the emergence of life (*e.g.*, Wolf and Toon 2010), and $O_2$ would have been lethal for the last universal common ancestor of life on Earth, an obligate anaerobe (Di Giulio 2007; Weiss et al 2016). In fact, recently proposed abiotic mechanisms for generating substantial amounts of $O_2$ on some exoplanets may preclude their habitability for this reason (see Meadows 2017 for review).

The abundant $O_2$ in Earth's atmosphere today, upon which all macroscopic life depends, is biologically produced by oxygenic photosynthesis. The origin of oxygenic photosynthesis was a critical prerequisite for the subsequent transition from a simple, anaerobic biosphere to today's complex, aerobic world, and this innovation is therefore among the most important events in Earth history. But Earth's oxygenation was protracted, and the relationship between biological $O_2$ production, oceanic oxygenation, and atmospheric $O_2$ accumulation is not straightforward (reviewed in Lyons et al 2014). In this section, we review geochemical proxy and model constraints on Earth's oxygenation trajectory (summarized in **Fig. 1**), and discuss links between this $O_2$ history and the timing and tempo of biological innovation (**Fig. 2**).

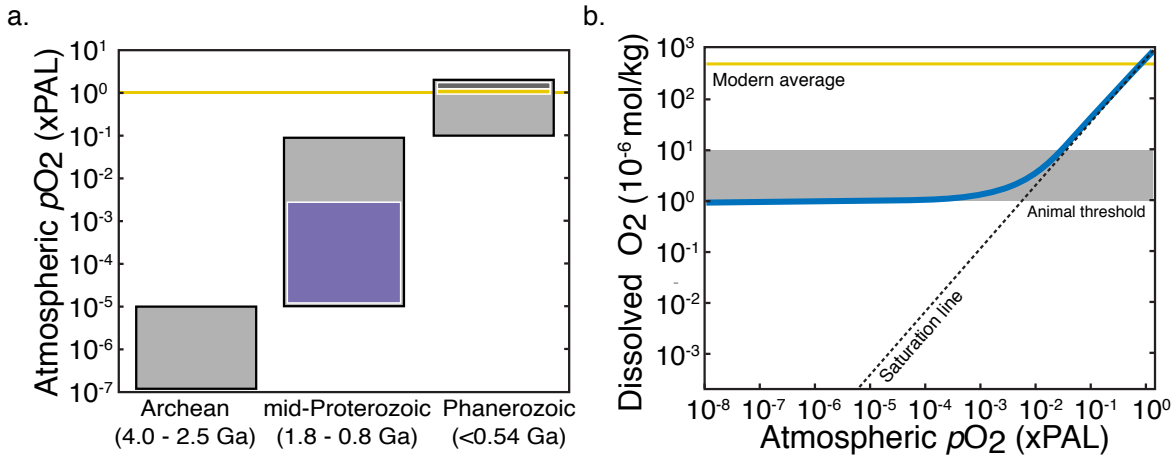

**Figure 1.** *Atmospheric and oceanic $O_2$ evolution.* In (**a**), the grey boxes represent inclusive ranges of model and proxy-based constraints on atmospheric $O_2$ for each geological eon. The minimum and maximum values for each grey box are specified in **Table 1**. The colored bars in (**a**) represent preferred ranges corresponding to constraints from specific archives discussed in the text, including: Cr isotopes (purple), and Mauna Loa observations (yellow). The dark grey box represents the *Carboniferous* Period (359-299 Ma; Berner 1999). The solid blue line in (**b**) shows the modeled relationship between atmospheric $pO_2$ and globally averaged dissolved $O_2$ in the shallow ocean (Olson et al 2013; Reinhard et al 2016). Also shown in (**b**) are calculated $O_2$ concentrations at thermodynamic equilibrium with respect to the sea-air exchange (dashed line) as well as estimates of early animal $O_2$ requirements (shading; D Mills et al 2014). The $O_2$ content of the deep ocean has scaled with the $O_2$ content of the atmosphere and surface ocean in the recent past, but the subsurface ocean remains strictly anoxic for $pO_2$ less than ~40% PAL based on the results of a simple box model (Canfield 1998; **Table 2**). The $O_2$ landscape of the ocean cannot be straightforwardly extrapolated from the $O_2$ content of the atmosphere—on Earth or on an exoplanet.

### *2.1. Oxygenic photosynthesis on the early Earth*

Oxygenic photosynthesis apparently emerged early in Earth's history (Buick 2008; Farquhar et al 2011). The biochemistry of oxygenic photosynthesis is complex (see Hohmann-Marriot & Blankenship 2012 for a review), but the net reaction is $CO_2 + H_2O + h\nu \rightarrow CH_2O + O_2$, where $h\nu$ represents photon energy from sunlight, and $CH_2O$ represents generalized organic matter. The exact timing of this biological innovation remains unclear, but geological evidence for oxygenic photosynthesis dates to at least ~3 Ga (Nisbet et al 2007; Planavsky et al 2014a)—significantly predating evidence for the initial accumulation of $O_2$ in Earth's atmosphere ~2.3-2.4 Ga during the 'Great Oxidation Event' (GOE; *e.g.*, Lyons et al 2014). Thus, oxygenic photosynthesis is, empirically, an insufficient prerequisite for atmospheric oxygenation on Earth-like planets.

Consequently, although large-scale $O_2$ production on Earth is exclusively biological, most models for atmospheric oxygenation during the GOE involve geologic controls that modulate $O_2$ consumption rather than net $O_2$ production. In other words, the GOE is generally regarded as a tipping point beyond which net biological production of $O_2$ irreversibly exceeded the collective geological sinks. Net $O_2$ production is controlled by burial of reduced organic C that escapes respiratory oxidation by photosynthetic $O_2$ (if all of the $O_2$ produced by photosynthesis were respired via the reverse reaction, $CH_2O + O_2 \rightarrow H_2O + CO_2$, there would be no net $O_2$ accumulation). In the Archean, strongly reducing volcanic outgassing may have suppressed $O_2$ accumulation despite widespread production at the Earth's surface, but this sink for photosynthetic $O_2$ may have declined as the thermal and tectonic state of the Earth evolved (Gaillard et al 2011; Kump and Barley 2007). Other models for Earth's delayed oxygenation emphasize the importance of long-term reductant loss via H escape to space, which results from the photolysis of H-bearing gases in the atmosphere and represents an irreversible oxidation of the planet that may have favorable implications for subsequent oxygenation (Catling et al 2001; Zahnle et al 2013).

Despite limited accumulation of $O_2$ in the Archean atmosphere, biogeochemical models predict the development of 'oxygen oases' (local to regional oxygenation) in highly productive areas of the surface ocean (Daines and Lenton 2016; Olson et al 2013) and in association with photosynthetic microbial mats (Herman and Kump, 2005; Lalonde and Konhauser 2015; Sumner et al 2015). Indeed, some geochemical proxies indicate the presence of dissolved $O_2$ in the shallow ocean beginning as early as ~3 Ga (Planavsky et al 2014a). Following these earliest hints of $O_2$, the geochemical record of Archean oxygen oases is increasingly robust in the ensuing >500-million-year lead up to the GOE (Czaja et al 2012; Duan et al 2010; Eigenbrode and Freeman 2006; Garvin et al 2009; Godfrey and Falkowski 2009; Kendall et al 2010; Kurzweil et al 2016; Riding et al 2014; Stüeken et al 2015a). It is unlikely that exoplanet analogs to these Archean oxygen oases, which lack atmospheric expression, would be remotely detectable (Reinhard et al 2017a), demonstrating that apparently anoxic exoplanetary environments may be highly productive and potentially capable of developing certain types of complex life (*e.g.*, D Mills et al 2014).

Some authors also report evidence for transient episodes of atmospheric oxygenation, or 'whiffs' of $O_2$, beginning at least 50 million years before the GOE based on the appearance of oxidative weathering of the continents (Anbar et al 2007; Reinhard et al 2009; Kendall et al 2015). The geochemical fingerprints of atmospheric whiffs of $O_2$ are readily distinguishable from oceanic oxygen oases (Reinhard et al 2013a), but, like oases, whiffs would not be recognizable in disk-averaged spectra of the distant Earth (Reinhard et al 2017a). Several authors have also questioned whether putative whiff signals may instead derive from oxidative transformations by microbial communities living within soils rather than oxidative weathering beneath an oxygenated atmosphere (Lalonde and Konhauser 2015; Sumner et al 2015). Clarifying the spatiotemporal dynamics of whiffs, and their relationship to oases, will require additional numerical modeling, but the existing data strongly suggest that $O_2$ was ecologically and geochemically important long before it was globally abundant (and thus remotely detectable) in the atmosphere (but see Fischer et al 2016).

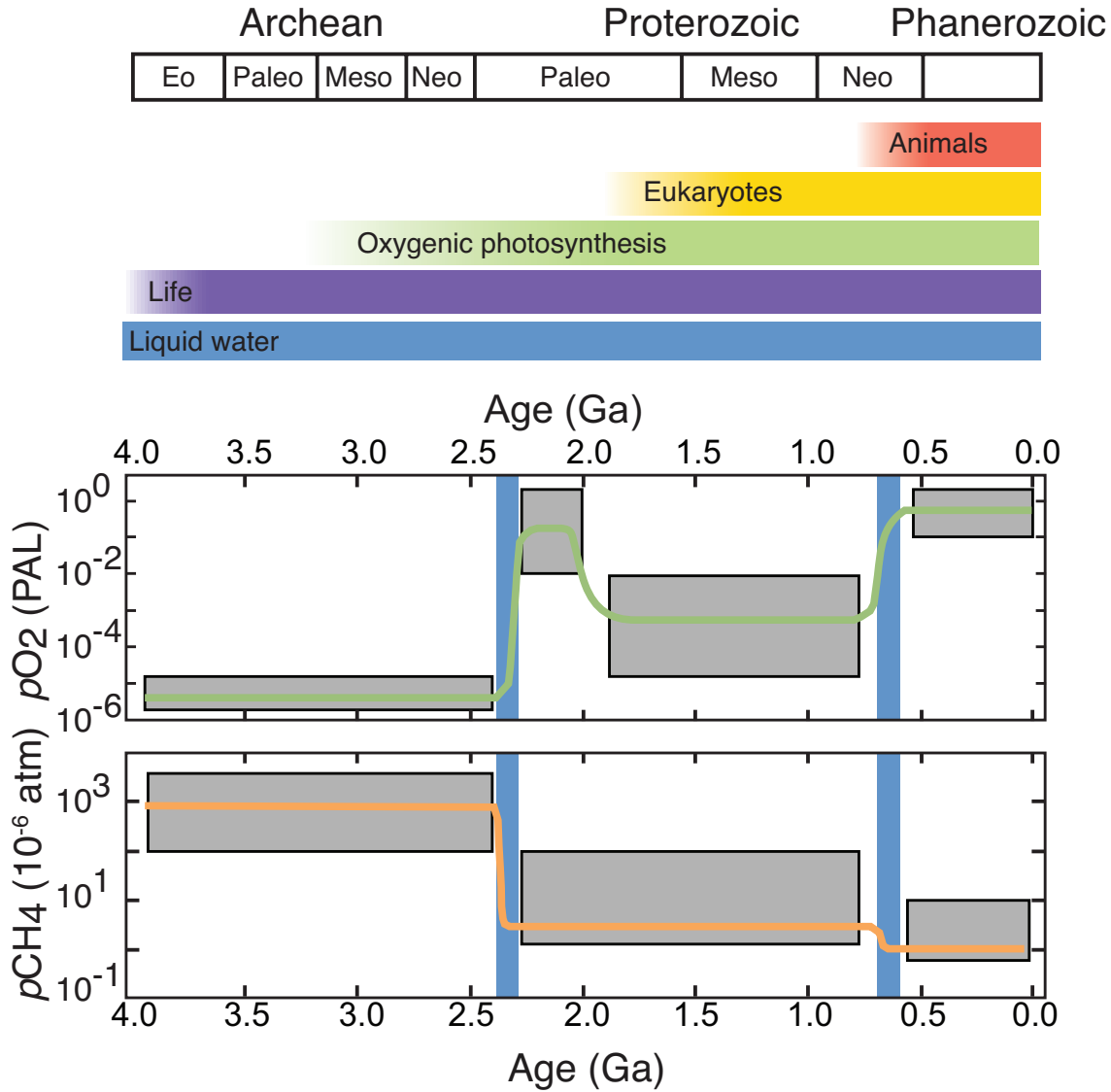

**Figure 2**. *Co-evolution of life and surface environments on Earth*. The top panel shows the timing of major transitions in the history of the biosphere. The middle panel shows Earth's oxygenation trajectory, while the bottom panel shows the abundance of $CH_4$ through time. In each, the vertical blue bars denote the timing of low-latitude glaciations, while colored lines show one possible trajectory through the parameter space implied by proxy reconstructions (shaded boxes; see **Fig. 1** and **Fig. 3b**).

## *2.2. The Great Oxidation Event*

The first of at least two major steps in oxygenation occurred at ~2.3-2.4 Ga during the GOE of the Paleoproterozoic, although other dramatic increases and decreases are likely (Lyons et al 2014). Numerous geochemical proxies capture this increase in $O_2$ levels. These proxies are diverse, and each has a unique sensitivity to environmental $O_2$, but all of the geochemical records broadly agree: the Earth system permanently changed during the GOE. Evidence for an increase in atmospheric $O_2$ includes the disappearance of detrital pyrite, which indicates a permanent and global onset of oxidative dissolution of $O_2$-sensitive pyrite during weathering (*e.g.*, Johnson et al 2014). Most compellingly, however, the GOE is marked by the loss of mass independent fractionation of sulfur isotopes (MIF-S) in marine sediments (Farquhar et al 2000). Large magnitude MIF-S is generated and preserved only when atmospheric $O_2$ is sufficiently low to preclude UV attenuation by $O_3$ and oxidative homogenization of photochemically produced S species, and the abrupt disappearance of this isotope signal at the GOE captures the rise of $O_2$ above $\sim 10^{-5}$ times the present atmospheric level (PAL; Pavlov and Kasting 2002).

The GOE is also broadly associated with the ~2.3 to 2.0 Ga Lomagundi 'event,' a large magnitude and long-lived positive C isotope excursion that is globally expressed in marine carbonates (Karhu and Holland 1996). Conventional interpretations of shifts towards heavy carbonates require enhanced removal of isotopically light organic C (*e.g.*, Kump and Arthur 1999). Assuming $O_2$-producing cyanobacteria were ecologically dominant, elevated organic burial implies a correspondingly large $O_2$ release to the atmosphere (*e.g.*, Karhu and Holland 1996)—potentially resulting in $O_2$ levels as high as ~12x the modern atmospheric $O_2$ inventory during the Lomagundi event by some estimates (Rybacki et al 2016). Indeed, several additional geochemical records suggest elevated atmospheric $pO_2$ at this time (Scott et al 2008; Planavsky et al 2012; Partin et al 2013; Hardisty et al 2014).

Despite major oxygenation during the GOE, Earth's oxygenation was neither stepwise nor unidirectional. It appears that $O_2$ levels fell precipitously in the wake of the Lomagundi event (*e.g.*, Planavsky et al 2012; Partin et al 2013). This 'oxygen overshoot' is not well understood. One possibility is that oxygenation was initially perpetuated by positive feedbacks involving acid weathering as the first large-scale oxidation of crustal sulfides by atmospheric $O_2$ generated sulfuric acid, enhancing nutrient fluxes to the ocean and favoring continued $O_2$ accumulation via organic carbon burial (Bekker and Holland 2012; Konhauser et al 2011). Then, eventual deoxygenation may have been driven by the exposure and oxidation of the reduced C that was sequestered during the Lomagundi event (Bekker and Holland 2012; Kump et al 2011). Even more enigmatically, atmospheric $O_2$ remained low during the ensuing mid-Proterozoic (~1.8 to 0.8 Ga; Cole et al 2016; Planavsky et al 2014b)—failing to achieve near modern levels, or even potentially remotely detectable levels (Reinhard et al 2017a), for much more than a billion years after the GOE and nearly 2.5 billion years after the origin of oxygenic photosynthesis.

### *2.3. Oxygen during Earth's middle chapter*
The mid-Proterozoic world has traditionally been envisaged as an intermediate state between the Archean and Phanerozoic, with $O_2$ levels that were a significant fraction of modern (*i.e.*, ~1-40% PAL; see Kump 2008). The upper limit is based on the atmospheric $pO_2$ conditions for which deep-ocean anoxia is maintained in a simple box model (Canfield 1998), and thus it carries considerable uncertainty embedded in unknowns regarding the marine biological pump at that time, among other factors. The lower limit reflects estimates of the $O_2$ levels that would be necessary to retain insoluble $Fe^{3+}$ during weathering following the GOE—in stark contrast with the Archean scenario, where Fe was mobile as $Fe^{2+}$ during weathering (*e.g.*, Rye and Holland 1998). These calculations based on paleosols (ancient soils), however, are imprecise and require independent constraints on $pCO_2$ in order to approximate $pO_2$. Given these uncertainties, $Fe^{3+}$ retention in paleosols is compatible with $O_2$ levels that are substantially lower than the frequently cited 1% threshold (Pinto and Holland 1988; Zbinden et al 1988). In combination with improved constraints suggesting lower levels of $CO_2$ during the Proterozoic (Kah and Riding 2007; Mitchell and Sheldon 2010; Sheldon 2006; see **Section 3**), the lower limit of ~1% PAL $O_2$ based on paleosols is almost certainly too high.

New isotopic proxies now suggest that $O_2$ levels were much lower than 1% PAL, and perhaps less than 0.1% PAL, until ~800 million years ago (Planavsky et al 2014b). This new upper limit, which is substantially lower than the tenuous lower limits based on paleosols, is obtained from the absence of chromium (Cr) isotope fractionation in Proterozoic marine sediments (Frei et al 2009; Planavsky et al 2014b; Cole et al 2016). Chromium isotope fractionation is imparted during redox transformations involving Mn oxides, which themselves require free $O_2$ for formation (Frei et al 2009 but see Johnson et al 2013 for a conflicting view); thus, the absence of Cr isotope fractionation speaks to the potential for exceptionally low $pO_2$ in the Proterozoic weathering environment despite the oxidative immobilization of Fe as insoluble $Fe^{3+}$.

**Table 1.** Atmospheric $O_2$ constraints for each geologic eon

| Eon | Constraints (xPAL) | | | Notes |
|---|---|---|---|---|
| | | *Min.* | *Max.* | |
| **Archean** | | $10^{-12}$ | $10^{-5}$ | The minimum estimate arises from abiotic photochemical production of $O_2$ (1); the maximum derives from the persistence of MIF-S (2), but transient excursions to higher $pO_2$ (3) are allowed (4). |
| **mid-Proterozoic** | *Incl.* | $10^{-5}$ | $10^{-1}$ | The minimum is constrained by absence of MIF-S (2); the maximum is likely constrained by the absence of Cr isotope fractionation (5), but is difficult to reconcile with photochemical models (6). |
| | *Pref.* | $10^{-5}$ | $10^{-3}$ | |
| **Phanerozoic** | | $10^{-1}$ | 2 | The minimum and maximum values here reflect temporal variability rather than ambiguities in proxy interpretation as above (7). |

*By convention, $pO_2$ is expressed with respect to the present atmospheric level (PAL) of $O_2$: 0.21 atm. Minimum and maximum values are provided for inclusive and preferred ranges where divergent constraints exist. Inclusive ranges correspond to the grey boxes in **Fig. 1** whereas preferred ranges are highlighted with colored boxes. The numbered references within the table correspond to: (1) Kasting et al 1979; (2) Pavlov and Kasting 2002; (3) Anbar et al 2007; (4) Reinhard et al 2013b; (5) Planavsky et al 2014b; (6) Claire et al 2006; (7) Berner 1999.*

Distinguishing between low and very low O$_2$ levels, although geochemically nuanced, is biologically significant. If atmospheric O$_2$ was <<1% modern levels in the Proterozoic, the ocean would have been characterized by a spatially and temporally patchy O$_2$ landscape that was more similar to the Archean ocean than conditions during the Phanerozoic (Olson et al 2013; Reinhard et al 2016; **Fig. 1b**). In this scenario, the shallow ocean would have been poorly buffered against nighttime and seasonal anoxia despite local accumulation of photosynthetic O$_2$ (Reinhard et al 2016). Indeed, the persistence of limited iodate (IO$_3^-$) incorporation into marine carbonates is best explained by spatially heterogeneous and/or temporally unstable ocean oxygenation and the possibility of frequent anoxia and even H$_2$S in the shallowest ocean (Hardisty et al 2017). Such an environment may explain the delayed diversification of eukaryotes after their early origin ~1.7-1.6 Ga (Gilleaudeau and Kah 2015; Knoll 2014; Knoll and Nowak 2017).

A patchy O$_2$ landscape in space and time, rather than uniformly low concentrations in the surface ocean, may also explain the absence of animals at this time (Planavsky et al 2014b; Reinhard et al 2016), given that the earliest animals would have required very low levels of O$_2$ (Butterfield 2009; D Mills et al 2014; Sperling et al 2013). It is also worth noting that at these low levels, production of O$_3$ from O$_2$ is impacted (Kasting and Donahue 1980). Ineffective UV filtering by O$_3$, and thus elevated fluxes of DNA-damaging UV at the Earth's surface, may have limited the complexity of terrestrial ecosystems throughout the Proterozoic while also having important implications for the photochemical stability of key greenhouse gases (Arney et al 2016; Olson et al 2016b; see **Section 4**).

Despite an increasingly complete view of Proterozoic O$_2$ dynamics through the lens of geochemical proxies, we still lack a mechanistic model for stabilizing atmospheric O$_2$ at the levels implied by the proxy records. In particular, <<1 % modern $p$O$_2$ is difficult to reconcile with photochemical models that predict small fluctuations in O$_2$ fluxes around this low baseline should result in either runaway oxygenation to ~10% modern or collapse to pre-GOE levels (Claire et al 2006; Goldblatt et al 2006; Zahnle et al 2006). Notably, however, these models lack dynamic boundary conditions; that is, they do not allow for changing O$_2$ fluxes in response to evolving saturation conditions or the possibility that biological O$_2$ production responds to ambient O$_2$—and thus these models preclude potential negative feedbacks that would tend to stabilize $p$O$_2$ (Olson et al 2016a). Indeed, several authors suggest mechanisms by which Proterozoic O$_2$ production may be modulated by O$_2$-induced micro- or macronutrient limitation (Anbar and Knoll 2002; Fennel et al 2005; Laakso and Schrag 2014; Olson et al 2016a; Reinhard et al 2013c, 2017b). Reconciling geochemical evidence for low atmospheric $p$O$_2$ with comprehensive models for stability of the O$_2$ cycle is a frontier for current research. Nevertheless, it is clear that atmospheric O$_2$ levels during Earth's middle age were distinct from those before and after in Earth history.

### *2.4. Neoproterozoic oxygen dynamics and the rise of animals*
Following this interval of apparent biogeochemical stasis marked by low O$_2$, it is widely believed that there was a Neoproterozoic Oxidation Event (NOE) analogous to the Paleoproterozoic GOE (*e.g.*, Campbell and Squire 2010; Canfield et al 2007; Sahoo et al 2012). Unlike the GOE, however, the timing and magnitude of oxygenation in the Neoproterozoic is poorly constrained (Och and Shields-Zhou 2012). This transition is traditionally depicted as a step increase in $O_2$ in the Neoproterozoic (see Fig. 2 from Kump 2008), but it likely occurred in several stages (Fike et al 2006). Beginning at ~0.8 Ga, Cr isotopes are persistently fractionated in marine sediments—suggesting a permanent, but modest, increase in atmospheric $O_2$ above the very low $pO_2$ threshold at which oxidative Cr cycling is enabled (Cole et al 2016; Planavsky et al 2014b). Meanwhile, there was a persistence of anoxic conditions in the deep ocean despite rising $O_2$ in the atmosphere (Canfield et al 2008; Dahl et al 2011; Johnston et al 2013; Li et al 2010; Sperling et al 2015b). Superimposed on these broad steps of oxygenation are suggestions of a dynamic redox environment in which transient pulses of oxygenation, analogous to pre-GOE whiffs, apparently punctuate a broadly anoxic background throughout the Neoproterozoic (Li et al 2015; McFadden et al 2008; Sahoo et al 2016; but see Lau et al 2016). A terminal oxygenation event in which near modern $pO_2$ was finally achieved has yet to be identified, but it now seems this transition likely occurred well into the Phanerozoic (<542 million years ago (Ma); Johnston et al 2013; Sperling et al 2015b)—perhaps as late as the Devonian (419-359 Ma) with the rise of land plants (Dahl et al 2010; Lenton et al 2016; Wallace et al 2017).

**Table 2.** Oceanic $O_2$ constraints for each geologic eon

| Eon | Constraints (µM) | | Notes |
|---|---|---|---|
| | *Min.* | *Max.* | |
| **Archean** | *Surf.* 0 | 10 | The Archean ocean was anoxic, with the possible exception of oxygen oases in productive regions of the shallow ocean following the origin of oxygenic photosynthesis (1). The maximum value here represents a local, rather than global, maximum. |
| | *Deep* 0 | 0 | |
| **mid-Proterozoic** | *Surf.* 0 | 25 | The shallow portions of the Proterozoic ocean were heterogeneously oxygenated, both spatially and temporally (2), while the deep oceans remained anoxic (3). The range of surface ocean values shown here brackets both spatial and temporal variability. |
| | *Deep* 0 | 0 | |
| **Phanerozoic** | *Surf.* 25 | 500 | The surface ocean range here is calculated based on the assumption of equilibrium sea-air exchange for $pO_2$ between 10-200% modern (**Table 1**). The $O_2$ content of the ocean dramatically increased in the Phanerozoic (4), but the redox landscape of the subsurface ocean is spatially and temporally variable—independent of atmospheric $pO_2$ trends (*e.g.*, 5). |
| | *Deep* 0 | 500 | |

*Unlike $O_2$ in the atmosphere, dissolved $O_2$ is strongly heterogeneous in the ocean, which mixes over much longer timescales. Whereas the surface ocean is a site of net $O_2$ production via photosynthesis, $O_2$ consumption by respiration dominates in the dark, subsurface ocean—further amplifying the potential for heterogeneity. The numbered references within the table correspond to: (1) Olson et al 2013; (2) Reinhard et al 2016; (3) Dahl et al 2011; (4) Dahl et al 2010; (5) Owens et al 2013.*

Meanwhile, although O$_2$ is accepted as a perquisite for animal life on Earth, the relationship between Earth's dynamic oxygenation history and the emergence of animals remains controversial (see Sperling et al 2015a). Nonetheless, there are several intriguing associations that are worth noting. First, a major diversification of eukaryotes is roughly coincident with the first hints of Neoproterozoic oxygenation at ~0.8 Ga (Knoll 2014; Planavsky et al 2014b). In the Ediacaran (645-541 Ma), episodes of microfossil diversification may also correlate with pulses of oxygenation (Li et al 2015; McFadden et al 2008). Morphological and behavioral complexity, both requiring relatively high levels of O$_2$, finally emerge in the late Ediacaran (*e.g.*, Droser and Gehling 2015), perhaps as long as 100 million years or more after the earliest molecular traces of simple animals with comparatively low O$_2$ requirements (Love et al 2009; D Mills et al 2014)—and coincident with yet another oxygenation event or series of events (Hardisty et al 2017; Li et al 2017; Sahoo et al 2016; Wallace et al 2017).

Although the data are not definitive regarding an O$_2$ threshold for the emergence of animals on Earth, the existing data do suggest that, at minimum, O$_2$ availability was an important constraint on the tempo and mode of their subsequent diversification. Thus, quantification of O$_2$ in exoplanetary atmospheres may provide constraints on the likelihood of biospheric progression towards complex multicellularity. However, we do not yet know (1) why atmospheric O$_2$ levels remained persistently low during the mid-Proterozoic or (2) how atmospheric O$_2$ eventually achieved modern levels. Until we fully understand whether pervasive oxygenation is an inevitable outcome, or even a likely consequence, of oxygenic photosynthesis, it is difficult to assess the likelihood of detecting O$_2$ as a biosignature in an exoplanet atmosphere (Gebaur et al 2017; Reinhard et al 2017a). Without a predictive model for planetary oxygenation, the likelihood of complex multicellularity and even intelligence beyond Earth also remains unconstrained (see Catling et al 2005).

### *2.5. Continued oxygen evolution in the Phanerozoic*

Following the achievement of near modern levels, O$_2$ has remained sufficiently high to support the emergence of increasingly diverse and complex life and ecologies. As for the Proterozoic, however, persistent oxygenation does not imply invariance; dynamic variability has continued throughout the Phanerozoic and has actually increased in absolute magnitude over time compared to the Precambrian. Most notably, the Carboniferous Period (359-299 Ma), named for its globally extensive coal deposits (*i.e.*, extensive organic C burial) and known for its prevalence of gigantism among insects (Graham et al 1995), is thought to have been a time of particularly high atmospheric *p*O$_2$, likely approaching ~0.4 atm (35% of the atmosphere after correcting for the increase in total pressure)—which is nearly double the modern level (Berner 1999). We acknowledge that the coarse spatial and temporal scales over which our sparse proxy record integrates likely mask similar fluctuations in Precambrian *p*O$_2$ over many orders of magnitude around the low baseline level, albeit of much smaller absolute magnitude compared to the shifts of the high *p*O$_2$ Phanerozoic. Put another way, even the small seasonal oscillation in atmospheric *p*O$_2$ that occurs today reflects a change in O$_2$ that is of similar magnitude to the entire O$_2$ content of the mid-Proterozoic atmosphere (Keeling and Shertz 1992; Planavsky et al 2014b). In addition to sorting out the enigmatic details of O$_2$ cycle stability during each geologic eon, future work constraining the magnitude, timescales, and biogeochemical implications of O$_2$ fluctuations within each eon will be important for remotely characterizing exoplanetary environments because atmospheric spectra will ultimately yield single snapshots with similar limitations as our geochemical proxy reconstructions.

## 3. Carbon dioxide, climate regulation, and enduring habitability

Despite similar starting conditions, Venus, Earth, and Mars have had very different climatic fates. The Earth has avoided succumbing to either an irreversible glaciation or a runaway greenhouse—despite episodes of extreme, low-latitude glaciation and intervals of dramatic warming against a backdrop of continuously increasing solar luminosity (*e.g.*, Evans et al 1997). This remarkable stability of Earth's climate, and thus Earth's persistent habitability, depends on long-term climate stabilization via feedbacks involving atmospheric $CO_2$ (Walker et al 1981). In this section, we examine the climate regulation mechanisms that operate on Earth—and likely on other habitable worlds (Kasting et al 1993). We also discuss the geological controls on atmospheric $CO_2$ levels, and we review existing proxy constraints on $CO_2$ through time (**Fig. 3a**).

### *3.1. The faint young Sun paradox*

The Sun was substantially dimmer when Earth formed (*e.g.*, Gough 1981). If Earth had always had its current atmosphere, Earth would have been extensively glaciated for at least the first ~2.5 billion years of its history (Sagan and Mullen 1972). Of course, this icy past is in conflict with evidence for liquid water shortly following the Moon-forming impact, as early as 4.4 Ga (Valley et al 2002), as well as a general absence of Archean glacial deposits—implying surface temperatures that were similar to or perhaps greater than today's. Indeed, geochemical proxies suggest the early Earth was modestly warmer (Blake et al 2010; Hren et al 2009). Several other authors have argued for a very hot Archean climate (Garcia et al 2017; Knauth and Lowe 2003; Robert and Chaussidon 2006), but the geochemical and biochemical evidence for hot, rather than warm, conditions remains controversial (Kasting et al 2006; Hren et al 2009). In either case, reconciling stellar evolution with an ancient climate that was warmer than today likely requires that the magnitude of greenhouse warming was much greater during Earth's early history (Newman and Rood 1977; Owen et al 1979; Sagan and Mullen 1972; Walker et al 1981). Recognizing that Earth's early atmosphere was reducing and lacked $O_2$, Sagan and Mullen (1972) suggested that enhanced greenhouse warming by ammonia ($NH_3$) could have compensated for reduced solar luminosity. This scenario, however, is problematic because $NH_3$ is photochemically unstable and lacks natural sources that are sufficient to sustain large atmospheric concentrations; instead, it is more likely that $CO_2$ levels were much higher because, unlike $NH_3$, geological sources (*e.g.*, volcanism) provide large fluxes of $CO_2$ to the atmosphere (Owen et al 1979; Kasting 1993; Walker et al 1981; Walker 1990).

### *3.2. The silicate weathering thermostat*

The warming potential of $CO_2$ is regulated by a negative feedback through the temperature dependence of silicate weathering kinetics (Walker et al 1981). Whereas low temperatures result in low weathering rates and limited $CO_2$ drawdown, high temperatures support high weathering rates and enhanced $CO_2$ drawdown.  In this classic scenario, low surface temperatures and sluggish weathering kinetics allow $CO_2$ levels to increase naturally to compensate for the faint young Sun, thus maintaining clement conditions on the early Earth. This feedback also provides a mechanism for Earth's recovery from low-latitude glaciation: continued release of volcanic $CO_2$ during the glacial event allows atmospheric $CO_2$ to build up to very high levels when extensive ice coverage suppresses subaerial silicate weathering (Hoffman et al 1998 but see Le Hir et al 2008). Conversely, high $CO_2$ in the wake of deglaciation would stimulate high weathering rates—ultimately drawing down $CO_2$ and re-stabilizing climate. This powerful feedback thus regulates Earth's surface temperature, forms the basis of our understanding of long-term habitability of Earth, and underlies the definition of exoplanetary habitable zones, which guide our search for life elsewhere (Kasting et al 1993; Kopparapu et al 2013).

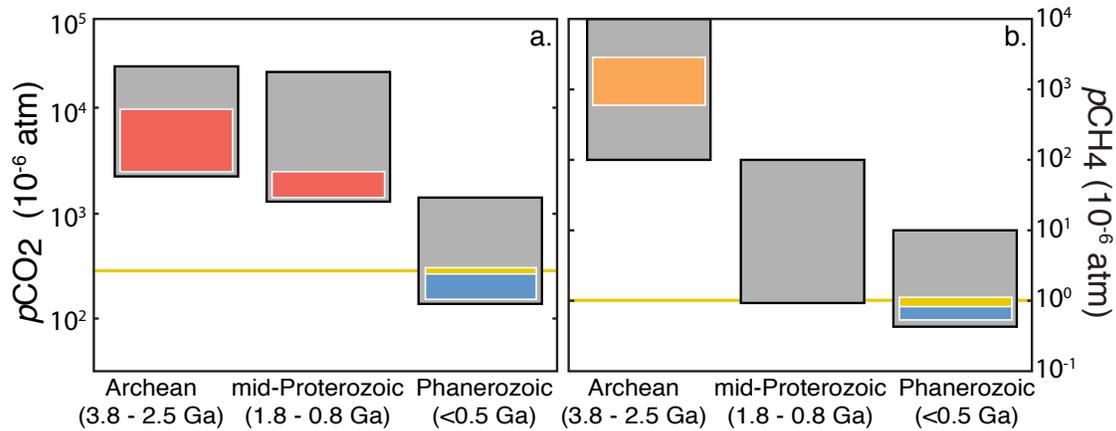

**Figure 3.** *Greenhouse constraints through Earth history.* For each geological eon, grey boxes represent inclusive ranges for model and proxy-based constraints on atmospheric $pCO_2$ (**a**) and $pCH_4$ (**b**). The minimum and maximum values for each grey box are specified in **Table 3** for $CO_2$ and **Table 4** for $CH_4$. The colored bars represent preferred ranges corresponding to constraints from specific proxies discussed in the text, including: paleosols (red), organic haze (orange), ice core records (for the last 800,000 years (Loulergue et al 2008; Luthi et al 2008); light blue), and Mauna Loa observations (since 1958 for $CO_2$ (*e.g.*, Keeling 1976) and since 1983 for $CH_4$ (Dlugokencky et al 1994); yellow).

There are, however, several complications that currently preclude confident prediction of ancient $CO_2$ levels strictly as a function of solar luminosity and volcanic outgassing rates (see Krissansen-Totton and Catling 2017). Although conceptually simple, the operation of the weathering thermostat has necessarily changed through time. On the earliest Earth, continental area was dramatically lower than today. This reduction of subaerially exposed silicates may impact the effectiveness of the coupling between temperature and $CO_2$ drawdown via continental weathering while also limiting the accumulation of carbon in shelf sediments (Lee et al 2016; Walker 1990). Meanwhile, the coupling between seafloor weathering and atmospheric $CO_2$ is poorly constrained (*e.g.*, Sleep and Zahnle 2001), with some models suggesting that $CO_2$ consumption during seafloor alteration may be more strongly controlled by Earth's thermal and tectonic evolution than Earth's atmospheric composition (Brady and Gislason 1997; Krissansen-Totton and Catling 2017). There are divergent models for the growth of the continents (reviewed by Cawood et al 2013), but most suggest significant continent formation in the late Archean (*e.g.*, Belousova et al 2010; Dhuime et al 2012)—and some model scenarios suggest that the areal extent of exposed continental crust continued increasing through the late Proterozoic (Condi and Aster 2010; Hawkesworth et al 2016). Furthermore, the weatherability of continental crust is unlikely to have been static throughout the development of terrestrial ecosystems (*e.g.*, Volk 1987). In particular, the impact of land plants, which are a recent evolutionary development (Kenrick and Crane 1997), must be considered (Lenton and Watson 2004). Thus, $CO_2$ drawdown via silicate weathering at a particular temperature has necessarily, but not straightforwardly, changed through time as the balance of seafloor and continental weathering has evolved and as Earth's subaerially exposed crust has become colonized by progressively more complex ecosystems (B Mills et al 2014).

### *3.3. Geological constraints on carbon dioxide through time*

Despite these complications, the expectation that ancient $pCO_2$ was higher in Earth's distant past than today is generally validated by geochemical proxies. Early analyses of Archean soils (paleosols) suggested $pCO_2$ less than ~40,000 μatm based on the absence of ferrous carbonate minerals (Rye et al 1995). Meanwhile, the formation of ferrous carbonate minerals in weathering rinds of 3.2 Ga Archean river gravels imply $CO_2$ levels greater than ~2,500 μatm (Hessler et al 2004). In sum, these data likely constrain Archean $pCO_2$ to ~10-140x modern pre-industrial levels of 280 μatm. More recently, updated calculations yield somewhat lower values, instead suggesting $pCO_2$ between ~10-50x pre-industrial levels and allowing for a declining trend through the Archean (Sheldon 2006; Driese et al 2011; for a conflicting view see Kanzaki and Murakami 2015).

The $CO_2$ levels permitted by these records, however, are much lower than what is required to compensate for reduced solar luminosity, suggesting that $CO_2$ was not sufficiently elevated to single handedly reconcile warm, apparently ice-free, conditions with the faint young Sun during the Archean (*e.g.*, Rye et al 1995). Inadequate warming by $CO_2$ does not imply that the silicate weathering feedback is not a powerful regulator of climate; instead, assuming that the proxy records have been correctly interpreted, the mismatch between Archean $pCO_2$ reconstructions and climate models indicates that there are other factors that positively influenced the Archean Earth's radiative budget. For example, higher atmospheric pressure may have amplified greenhouse warming via pressure broadening (Goldblatt et al 2009), but this scenario is incompatible with proxy records that suggest lower total atmospheric pressure in the Archean (Marty et al 2013; Som et al 2016; see **Section 5**). Alternatively, reduced continental area on the early Earth may have resulted in lower planetary albedo, which would have helped to warm the early Earth by limiting the reflection of solar energy (Rosing et al 2010). However, even complete removal of today's reflective land mass is likely to be insufficient to maintain temperatures similar to or greater than modern (Goldblatt and Zahnle 2011b). Differences in cloud structure and coverage may also modestly contribute to reduced planetary albedo, but reduced cloud reflectivity is also unlikely to completely compensate for reduced solar luminosity (Goldblatt and Zahnle 2011a). It seems the most promising scenarios for Archean warmth require either a combination of many small influences or much higher abundances of other greenhouse gases—particularly $CH_4$ (Kasting 2005; Kharecha et al 2005; Pavlov et al 2000, 2001; see **Section 4**).

In the Proterozoic, there are also geological indications of modestly elevated $pCO_2$. The morphological complexity of Proterozoic microfossils suggests $pCO_2$ of 10-200x PAL (Kaufman and Xiao 2003), but the fossil record also suggests that Proterozoic cyanobacteria utilized carbon-concentration mechanisms to fix $CO_2$ (Kah and Riding 2007). The need to concentrate $CO_2$ intracellularly probably limits environmental $pCO_2$ to less than 7-10x pre-industrial levels (Kah and Riding 2007), in good agreement with the paleosol records (Mitchell and Sheldon 2010; Sheldon 2006, 2013).

**Table 3.** Atmospheric $CO_2$ constraints for each geologic eon

| Eon | Constraints (µatm) | | Notes |
|---|---|---|---|
| | *Min.* | *Max.* | |
| **Archean** | *Incl.* 2500 | 40000 | The inclusive minimum and maximum constraints come from river gravels and paleosols, respectively (1, 2). The likely maximum reflects a refined paleosol constraint using updated methodology (3). The range in $pCO_2$ results from ambiguities in proxy records as well as secular decline (4). |
| | *Pref.* 2500 | 15000 | |
| **mid-Proterozoic** | *Incl.* 1400 | 28000 | The minimum value reflects a minimum reported upper estimate rather than a true lower bound (5); the inclusive maximum value is inferred from microfossil morphology (6) whereas the likely maximum is derived from paleosols (7). The range results from ambiguities in proxy records as well as secular decline. |
| | *Pref.* 1400 | 2800 | |
| **Phanerozoic** | 200 | 2800 | The Phanerozoic $CO_2$ history is well constrained (8); the range of values presented here reflects temporal variability in $pCO_2$. Despite a nonlinear trajectory, this record is broadly compatible with secular decline since the Archean (4). |

*Here $pCO_2$ is expressed in units of uatm as plotted in **Fig 3**, whereas paleo-$pCO_2$ constraints are often expressed as a multiple of the pre-industrial atmospheric level (PAL) in the Precambrian literature and/or ppmv in the more recent past. We have converted to µatm from PAL assuming $pCO_2$ = 280 µatm, unless otherwise specified by the original authors. Note that in the recent past for which total pressure has been 1 atm, 1 µatm is synonymous with 1 ppmv—but this equivalence is invalid for most of Earth history because total atmospheric pressure has changed substantially (see **Fig. 4b**) and we have thus avoided use of ppmv here. Minimum and maximum values are provided for inclusive and preferred ranges where divergent constraints exist. Inclusive ranges correspond to the grey boxes in **Fig. 3a** whereas preferred ranges are highlighted with colored boxes. The numbered references within the table refer to: (1) Hessler et al 2004 (2) Rye et al 1995; (3) Sheldon 2006; (4) Walker et al 1981; (5) Mitchell and Sheldon 2010; (6) Kaufman and Xiao 2003; (7) Sheldon 2013; (8) Royer et al 2004.*

As during the Archean, the $CO_2$ levels permitted by these records are probably insufficient to compensate for a less luminous sun without enhanced contribution from another greenhouse gas. Again, $CH_4$ is historically the most popular candidate (Catling et al 2002; Pavlov et al 2003; Fiorella and Sheldon 2017), but whereas abundant $CH_4$ is likely during the Archean (*e.g.*, Kharecha et al 2005), a $CH_4$ greenhouse is in conflict with several recent models of the Proterozoic $CH_4$ cycle (Daines and Lenton 2016; Laakso and Schrag 2017; Olson et al 2016b). $N_2O$ is also a possible warming agent at this time (Buick 2007; Roberson et al 2011), but elevated $N_2O$ is incompatible with some existing constraints on atmospheric $O_2$ due to the lack of associated photochemical shielding effects (Planavsky et al 2014b). Thus, despite significant progress, the faint young Sun paradox is not yet fully resolved for a broad swath of Earth history—and the mid-Proterozoic offers particular challenges. See **Section 4** for further discussion of $CH_4$ and its role in the Precambrian climate system.

In the Phanerozoic, a diverse collection of proxy records constrain atmospheric $CO_2$, including paleosols (*e.g.*, Cerling 1991), isotopic records (e.g., Freeman and Hayes 1992), leaf stomatal distributions (Van der Burgh et al 1993; McElwain and Chaloner 1995), and ice cores (Luthi et al 2008). These records are generally in good agreement with results from well-established carbon cycle models, including GEOCARB (*e.g.*, Berner and Kothavala 2001) and COPSE (Bergman et al 2004). In combination, the proxies and models suggest that $pCO_2$ has fluctuated within a factor of ~10 during the Phanerozoic (see compilation by Royer et al 2004). Although Phanerozoic $pCO_2$ evolution has not been unidirectional, modern (pre-industrial) $CO_2$ levels are among the lowest in Earth history—a result that is consistent with the expectation that $pCO_2$ has broadly declined as solar luminosity has unidirectionally increased. Given several significant deviations from an idealized decreasing trajectory, however, it is clear that evolving solar luminosity is not the only important lever on atmospheric $pCO_2$. As discussed above, geophysical and tectonic influences (*e.g.*, continental area, paleogeography, seafloor spreading rate, continental collision; see Raymo and Ruddiman 1992) and biological innovation (*e.g.*, the rise of land plants) are significant factors that may have profound impacts on atmospheric $pCO_2$ and Earth's climate dynamics. Thus, refining our understanding of the complexities of climate regulation on Earth—including the important roles of continent formation and exposure above the seas, mountain building, volcanism, and other first-order tectonic process—is integral to recognizing the diversity of worlds on which negative feedbacks provide long-term climate stability and delineating the distribution of habitable environments in the Universe.

## 4. A hazy role for methane in Earth's climate system
Methane is currently a trace constituent (~1 μatm) of Earth's atmosphere—but despite its low concentration, $CH_4$ is a critical component of the climate system. The atmospheric abundance and greenhouse contribution of $CH_4$ has changed significantly through time (*e.g.*, Kasting 2005). In this section, we review (1) the controls on $CH_4$ fluxes to the atmosphere, (2) model and geological constraints on atmospheric $CH_4$ levels through time (**Fig. 3b**), and (3) the role of atmospheric $CH_4$ in climate modulation and long-term maintenance of habitability on Earth. These controls, while discussed in an Earth context, could be universally relevant.

### *4.1. Methane as a climate savior in the Archean*

As discussed in **Section 3**, warming by $CH_4$ is widely invoked to reconcile existing $CO_2$ constraints with ice-free conditions on the early Earth, particularly in the Archean. Indeed, elevated $CH_4$ during the Archean is an attractive solution to the faint young Sun paradox because:

(1) biological $CH_4$ production, methananogenesis, is an ancient metabolism (Ueno et al 2006), and an anaerobic Archean biosphere may have had a high potential for $CH_4$ production (Kharecha et al 2005),
(2) abiotic sources of $CH_4$ (*e.g.*, volcanism, serpentinization, comets) may have been large early in Earth history, despite representing only 0.4% of the modern $CH_4$ source to the atmosphere (Emmanuel and Ague 2007),
(3) climatically significant $CH_4$ accumulation is possible, even for modest $CH_4$ fluxes, in the absence of atmospheric $O_2$ (Pavlov et al 2000), and
(4) the collapse of $CH_4$ may provide a mechanistic link between oxidation and glaciation as seen during the Paleoproterozoic (Pavlov et al 2000, 2001; Zahnle et al 2006).

Several models attempt to place lower and upper limits on Archean $CH_4$. Pavlov et al (2000) estimated a lower limit of 100 µatm by calculating the $CH_4$ level that would be necessary to reconcile clement conditions with independent constraints on Archean $pCO_2$ (**Section 3**). It has since become clear that this early calculation overestimated the potential warming from $CH_4$ and thus underestimated the amount of $CH_4$ necessary (Byrne and Goldblatt 2015; Haqq-Misra et al 2008). An alternative approach for estimating Archean $pCH_4$ is to calculate the $CH_4$ levels that arise from a reasonable $CH_4$ flux to the atmosphere given independent constraints on $pO_2$. If the $CH_4$ flux to the Archean atmosphere was similar to the modern $CH_4$ flux (Kharecha et al 2005), low $O_2$ conditions in the atmosphere would have permitted $pCH_4$ similar to 1,000 µatm in the Archean (Pavlov et al 2000, 2001), compared to ~1 µatm in the pre-industrial atmosphere. If $CH_4$ fluxes were modestly higher than today, $CH_4$ levels up to 35,000 µatm may be possible (Kharecha et al 2005).

Despite these expectations for high $CH_4$ levels, it has been challenging to confirm suspicions of elevated atmospheric $CH_4$ because conventional proxies for $CH_4$ are sensitive to local $CH_4$ oxidation within the ocean rather than global $CH_4$ accumulation in the atmosphere. For example, isotopically light carbonates or organic C may indicate extensive biological oxidation or assimilation of $CH_4$, respectively, within marine sediments because $CH_4$ is strongly depleted in heavy $^{13}C$ relative to other C substrates (*e.g.*, Eigenbrode and Freeman 2006; Williford et al 2016). Although these proxies allow us to confidently identify signals of $CH_4$ recycling in sedimentary archives, these proxies ultimately provide very limited insight to the global abundance of $CH_4$ in the atmosphere.

Recently, however, anomalous S isotope fractionation may finally provide indirect constraints on atmospheric $CH_4$ levels (*e.g.*, Izon et al 2017). There are striking variations in the structure of the MIF-S record in the several hundred million years prior to the abrupt disappearance of MIF-S from the sedimentary archive during the GOE (Izon et al 2015, 2017; Kurzweil et al 2013; Thomazo et al 2009, 2013; Zerkle et al 2012). Some of the variability immediately prior to the GOE may be attributable to whiffs of $O_2$ that could impact both the production and preservation of MIF-S (Reinhard et al 2009), but the majority of the fluctuations are not readily attributable to increases in atmospheric $O_2$. Instead, some S isotope deviations may signify attenuation of UV radiation by an organic haze (Izon et al 2015, 2017; Kurzweil et al 2013; Thomazo et al 2009, 2013; Zerkle et al 2012), similar to that of Saturn's icy moon, Titan (*e.g.*, Trainer et al 2006).

Polymerization of atmospheric $CH_4$ to form an organic haze requires atmospheric $CH_4:CO_2$ ratios in excess of ~0.2 (Trainer et al 2006). If $pCO_2$ can be independently constrained by the Archean paleosol record, it is possible to estimate a lower limit for $pCH_4$ sufficient to prompt haze formation. If the lower estimates of Archean $pCO_2$ (~10x PAL) are correct, the existence of a haze suggests $CH_4$ levels of at least ~600 µatm; if $pCO_2$ was higher (~50x PAL), correspondingly higher $CH_4$ levels of ~3,000 µatm are implied (see Izon et al 2017 for a similar calculation). In either scenario, the existence of a haze is consistent with $CH_4$ fluxes similar to modern (Kharecha et al 2005).

**Table 4.** Atmospheric $CH_4$ constraints for each geologic eon

| Eon | Constraints (µatm) | | Notes |
|---|---|---|---|
| | *Min.* | *Max.* | |
| **Archean** | *Incl.* 100 | 35000 | The inclusive range is based on early calculations of the minimum levels of $CH_4$ necessary to compensate for the FYS (1) and the maximum $pCH_4$ resulting from reasonable biological fluxes (2). The preferred range is updated to reflect isotopic evidence for biologically modulated organic haze that implies a $CH_4:CO_2$ ratio near 0.2 (3). |
| | *Pref.* 600 | 3000 | |
| **mid-Proterozoic** | *Incl.* 1 | 100 | The difference between the inclusive and likely maximum values here are the result of differing assumptions regarding the efficiency of $CH_4$ oxidation by the marine biosphere in models of the $CH_4$ cycle (4,5). |
| | *Pref.* 1 | 10 | |
| **Phanerozoic** | 0.4 | 10 | The Phanerozoic $CH_4$ history is relatively well constrained; the range of values presented here reflects modeled temporal variability in $pCH_4$ (6). |

*Here $pCH_4$ is expressed in units of µatm as plotted in **Fig 3b**, whereas paleo-$pCH_4$ constraints are often expressed in units of ppmv. Although in 1 µatm is equivalent to 1 ppmv when total pressure is 1atm, we avoided the use of ppmv here given the likelihood that total atmospheric pressure has changed significantly throughout Earth history (see **Fig. 4b**). Minimum and maximum values are provided for inclusive and preferred ranges where divergent constraints exist. Inclusive ranges correspond to the grey boxes in **Fig. 3b** whereas preferred ranges are highlighted with colored boxes. The numbered references within the table refer to: (1) Izon et al 2017; (2) Kharecha et al 2005; (3) Pavlov et al 2001; (4) Pavlov et al 2003; (5) Olson et al 2016b; (6) Beerling et al 2009.*

Isotopic evidence for an organic haze may also provide an upper limit on Archean $p$CH$_4$. At increasingly elevated CH$_4$:CO$_2$ ratios, the resulting haze becomes increasingly thick. The maximum thickness of the haze is ultimately limited because organic hazes attenuate the UV radiation that is required for hydrocarbon polymerization—and are thus photochemically self-shielding (Arney et al 2016). The thickness of the haze may be further limited by negative biogeochemical feedbacks because as the haze becomes optically thick, further increases in CH$_4$ have a net cooling, rather than warming, effect on the Earth and methanogenesis is positively correlated with temperature (Domagal-Goldman et al 2008; Pavlov et al 2001). If the haze is sustained by biogenic CH$_4$, the abundance of CH$_4$, the thickness of the haze, and global temperatures are therefore regulated by a negative feedback in which very high levels of atmospheric CH$_4$ disfavor high levels of CH$_4$ production (Domagal-Goldman et al 2008; Pavlov et al 2001). If the source of atmospheric CH$_4$ was dominantly geological rather than biological (*e.g.*, from seafloor serpentinization), and thus insensitive to surface temperature, the persistence of a haze anti-greenhouse over geological timescales would encourage an increase in CO$_2$ as the result of lower temperatures and reduced silicate weathering rates—thus stabilizing the CH$_4$:CO$_2$ ratio, haze thickness, and global temperatures independent of CH$_4$ levels (Domagal-Goldman et al 2008; Pavlov et al 2001). Geochemical records, however, suggest episodic rather than continuous haze formation in the late Archean, implying a biological rather than geological control on haze development (Izon et al 2017; Thomazo 2009, 2013; Zerkle et al 2012). This distinction is important because, given the feedback between temperature and CH$_4$ production, a biologically modulated haze is likely incompatible with CH$_4$ in significant excess of ~1,000 µatm on long-term average assuming reasonable estimates of late Archean $p$CO$_2$.

Intriguingly, the most recent generation of general circulation models (GCMs) can achieve warm Archean climates at CH$_4$ levels of ~1,000 µatm assuming $p$CO$_2$ near the upper limit of paleosol constraints (Byrne and Goldblatt 2015; Charnay et al 2013; Wolf and Toon 2013). It is encouraging that the CH$_4$ levels that could explain both warm and hazy conditions are achievable without invoking extreme CH$_4$ fluxes (Izon et al 2017). Thus, it seems, at least for now, that disparate lines of reasoning have converged to constrain late Archean $p$CH$_4$ to several hundred to a few thousand µatm, while resolving the long-standing faint young Sun paradox (but see Laakso and Schrag 2017 for a contrasting view).

Although these relatively high levels of CH$_4$ are incompatible with significant accumulation atmospheric O$_2$, the high $p$CH$_4$ implied by a haze would favor H escape via CH$_4$ photolysis and thus irreversible planetary oxidation on geologic timescales with potential significance for Earth's oxygenation trajectory (Catling et al 2001; Izon et al 2017; Zahnle et al 2013). Meanwhile, high levels of CH$_4$ and the spectral fingerprints of a haze may also provide a remotely detectable biosignature for inhabited worlds lacking O$_2$ (Arney et al 2016, 2017).

### *4.2. Muted methane in the Proterozoic*
Methane dynamics in the aftermath of the GOE and throughout the Proterozoic are more problematic than those for earlier time periods. The Paleoproterozoic saw severe, low-latitude glaciation (*e.g.*, Evans et al 1997). These glacial events are widely attributed to the loss of warming by CH$_4$ associated with the GOE, but it is not clear if atmospheric CH$_4$ collapsed as a consequence of oxygenation (Kopp et al 2005; Pavlov et al 2000, 2001)—or if declining $pCH_4$ was itself the oxygenation trigger (Konhauser et al 2009; Zahnle et al 2006). In either case, $CH_4$ levels may have recovered following the GOE and the associated glaciations, owing to enhanced UV shielding by $O_3$ (a photochemical product of $O_2$), which would have resulted in a greater atmospheric lifetime for $CH_4$ despite the greater abundance of $O_2$ (Claire et al 2006; Goldblatt et al 2006). A greenhouse bolstered by $CH_4$, therefore, may explain the lack of mid-Proterozoic glacial depositions—at any latitude (Evans et al 1997)—until more than a billion years later when $O_2$ appears to rise again in the Neoproterozoic (Catling et al 2002; Pavlov et al 2003). This scenario, however, demands $O_2$ levels that may be in conflict with recent $pO_2$ proxy records (Planavsky et al 2014b). If mid-Proterozoic atmospheric $pO_2$ was more modest than previously assumed (see **Section 2**), the atmospheric lifetime of $CH_4$ would not be extended by $O_3$, and $pCH_4$ may have failed to recover following the GOE (Olson et al 2016b).

Independent of potential complications arising from low levels of atmospheric $O_2$ and ineffective $O_3$ shielding, persistently elevated $CH_4$ during the Proterozoic is also difficult to reconcile with the evolution of the marine biosphere. Pavlov et al (2003) calculated that $CH_4$ levels of ~100-300 µatm are possible during Proterozoic time, but their calculations assumed complete inhibition of methanotrophy as the result of limited oxidant (*e.g.*, $O_2$, $SO_4^{2-}$) availability in the Proterozoic ocean. They estimated that the absence of methanotrophy would allow $CH_4$ fluxes to the atmosphere that exceeded the modern flux by more than a factor of 10. Although $O_2$ would have been restricted to the surface ocean at this time (Reinhard et al 2016), this assumption of negligible $CH_4$ oxidation by $O_2$ is invalid because oxygenated surface waters provide an effective barrier to the exchange of $CH_4$ from the deep ocean to the atmosphere (Daines and Lenton 2016). Furthermore, anaerobic methanotrophy coupled to $SO_4^{2-}$ reduction efficiently destroys $CH_4$ at $SO_4^{2-}$ concentrations that are much lower than those reconstructed for the Proterozoic ocean (Beal et al 2011; Kah et al 2004). Methane destruction may also be coupled to the reduction of $Fe^{3+}$ (Beal et al 2009; Crowe et al 2011), which, unlike $SO_4^{2-}$, can be abundant even in the total absence of $O_2$ in the ocean-atmosphere system (*e.g.*, Konhauser 2002). Indeed, modern anoxic basins are a trivial source of $CH_4$ to the atmosphere despite substantial $CH_4$ production in sediments as the result of anaerobic $CH_4$ recycling (*e.g.*, Crowe et al 2011). Considering that the terrestrial environments that produce the greatest amounts of $CH_4$ today did not yet exist in the Proterozoic (but see Zhao et al 2017), it is reasonable to expect that the *net* biological source of $CH_4$ to the Proterozoic atmosphere might have been similar to or lower than modern, despite widespread marine anoxia and the likelihood of greater $CH_4$ production within the ocean (Bjerrum and Canfield 2011; Olson et al 2016b).

Perhaps then it is unsurprising that with each successive generation of improved biogeochemical and photochemical models, estimates of Proterozoic $pCH_4$ have unidirectionally declined, despite substantial differences in the construction and biases of each model. The most recent of these models suggest that $pCH_4$ was <10 µatm beyond the hazy Archean (Daines and Lenton 2016; Laakso and Schrag 2017; Olson et al 2016b)—and fully within the range of $pCH_4$ reconstructed for the Phanerozoic (Bartdorff et al 2008; Beerling et al 2009). Meanwhile, the inferred existence of an organic haze in the late Archean is the only evidence from the rock record for significantly elevated atmospheric $pCH_4$ at any point in Earth history. This paradigm shift from high $CH_4$ towards low $CH_4$ for the second half of Earth history requires reevaluation of the mechanisms responsible for ice-free conditions despite a fainter Sun during Proterozoic time. Low baseline $p$CH$_4$ during the Proterozoic also calls into question existing models for oxidative CH$_4$ collapse as a driver for low-latitude glaciation in the Neoproterozoic (Olson et al 2016b; but see Laakso and Schrag (2017) for an alternate scenario). Most important to the purposes of this review, however, the absence of either significant CH$_4$ or O$_2$ during much of the Proterozoic also leaves us without an obvious atmospheric expression of planetary inhabitation (biosignature) for at least a billion years of Earth history and highlights the likelihood of false negatives in our search for life elsewhere (Reinhard et al 2017a).

### *4.3. Phanerozoic climate perturbations: methane as a double agent*

With the eventual pervasive oxygenation of the ocean-atmosphere system, methanogenesis was relegated to increasingly rare oxidant-deficient marine environments. Methanogenesis occurs in reducing marine sediments today, but CH$_4$ produced in these environments does not readily evade anaerobic oxidation by SO$_4^{2-}$ within sediment pore waters (*e.g.*, Reeburgh 2007), not unlike the Proterozoic scenario. Today and in the more recent past, biogenic CH$_4$ may be further isolated from the atmosphere by hundreds to thousands of meters of oxygenated seawater, severely limiting the CH$_4$ flux that evades oxidation in the ocean. Thus, excluding anthropogenic emissions, anaerobic terrestrial ecosystems (*e.g.*, wetlands) have dominated the biogenic CH$_4$ flux to the atmosphere during Phanerozoic time. Following the rise of land plants, model reconstructions suggest that $p$CH$_4$ was similar to modern, less than a few µatm, during much of the Phanerozoic (Bartdorff et al 2008; Beerling et al 2009), while ice core records reveal that $p$CH$_4$ has been less than 1 µatm for the last 800,000 years (Loulergue et al 2008).

Although steady state $p$CH$_4$ levels have been generally low and unremarkable since the GOE, transient pulses of CH$_4$ accumulation are likely and may be climatically significant. Whereas the climatic consequences of atmospheric CH$_4$ were modulated by a negative haze feedback in the Archean (Domagal-Goldman et al 2008; Pavlov et al 2001), under haze-free, Phanerozoic conditions CH$_4$ may participate in positive, destabilizing feedbacks in which CH$_4$-induced warming triggers greater CH$_4$ release to the atmosphere (*e.g.*, Bjerrum and Canfield 2011). Furthermore, because the atmospheric CH$_4$ inventory, unlike $p$CO$_2$, is not buffered by steady-state exchange with the ocean, very rapid changes in the abundance of atmospheric CH$_4$ are possible (*e.g.*, Schrag et al 2002). Meanwhile, because atmospheric CH$_4$ levels have been low during Phanerozoic time (Bartdorff et al 2008; Beerling et al 2009), relatively large swings in temperature are possible for even modest changes in $p$CH$_4$— particularly compared to the high CH$_4$ Archean scenario where strong short wave absorption limits warming by CH$_4$ at high concentrations (Byrne and Goldblatt 2015). Thus, in contrast with its role as a climate stabilizer in the Precambrian, CH$_4$ is typically cast as an environmental disruptor during Phanerozoic time. For example, transient pulses of CH$_4$ to the atmosphere are widely invoked to explain biological and climatic perturbation in the Phanerozoic, including mass extinctions and hyperthermal events (*e.g.*, Knoll et al 2007; Pancost et al 2007). The role of CH$_4$ in non-steady state climate dynamics remains an active area of research, and—given the stark contrast between CH$_4$ cycling in the first and second halves of Earth history—future investigations will need to elucidate the potential climatic role of CH$_4$ and its prospects as a biosignature under diverse exoplanetary conditions.

## 5. Nitrogen: Earth's climate system under pressure

Dinitrogen is not chemically reactive in the atmosphere. Even though N is a bioessential element, triply bonded $N_2$ is also metabolically inaccessible to most life on Earth. Nonetheless, $N_2$ is a critical component of Earth's habitability. Earth's dense $N_2$ atmosphere affects the warming potential of $CO_2$ and $CH_4$ via pressure broadening of their absorption bands, and thus contributes to the maintenance of atmospheric pressure and temperature conditions that are conducive to the persistence of liquid water at the Earth's surface.

Yet the history of atmospheric $N_2$ remains poorly constrained (*e.g.*, Wordsworth 2016). There are two primary reasons that significant uncertainties exist: (1) the possibility that $pN_2$ may have changed substantially over Earth's history was only recently recognized (Goldblatt et al 2009; Som et al 2016) and (2) reconstruction of $pN_2$, which is effectively inert, is inherently challenging. Thus the magnitude, mechanisms, and consequences of this evolution are still under active investigation. In this section, we briefly review the existing constraints on atmospheric $pN_2$ and total atmospheric pressure throughout Earth history (**Fig. 4**) and discuss the processes that may drive long-term changes in $pN_2$. We also highlight remaining questions and future research directions.

If $N_2$ was initially outgassed to the atmosphere following the Moon-forming impact, $N_2$ levels may have been high on the early Earth, perhaps 2-3x present atmospheric levels based on the apparent N content of the crust (Goldblatt et al 2009; Wordsworth and Pierrehumbert 2013). Proxy constraints for $pN_2$ are limited, but Marty et al. (2013) used N isotopes to constrain early Archean (3.5-3.0 Ga) $pN_2$ to 0.52-1.1 atm compared to 0.78 atm today. Thus, within the uncertainty of their study, ancient $pN_2$ may have been either modestly higher or lower than today.

Other approaches attempt to diagnose total atmospheric pressure rather than $pN_2$ specifically. If atmospheric $pN_2$ was unchanged through time, we would expect reduced Archean atmospheric pressure in the face of low $O_2$ and, correspondingly, a ~15-20% increase in atmospheric pressure over time arising from the long-term accumulation of $O_2$ and drawdown of $CO_2$. Any evidence that suggests atmospheric pressure that is similar to or greater than modern on the early Earth, therefore, implies substantial changes in $pN_2$ even allowing for much higher $pCO_2$. Alternatively, changes in $pN_2$ may also be implied if lower atmospheric pressures are observed—but the net change must be greater than that attributable to low $O_2$ alone.

One paleobarometric technique for constraining total atmospheric pressure involves fossil raindrop impressions (Som et al 2012). Based on the size distribution of raindrop craters, Som et al (2012) constrained total atmospheric pressure to less than 2 times modern, but likely similar to or lower than modern. Kavanagh and Goldblatt (2015) subsequently demonstrated that raindrop crater statistics are very sensitive to rainfall rate and are ultimately a poor reflection of overlying atmospheric pressure—suggesting that the Som et al (2012) data instead constrained atmospheric pressure to less than ~10 times modern levels. Raindrop impressions, therefore, cannot provide reasonable limits on ancient atmospheric $pN_2$ because the $N_2$ levels permitted by this proxy are similar to the combined N content of the atmospheric, oceanic, and crustal reservoirs today (Johnson and Goldblatt 2015).

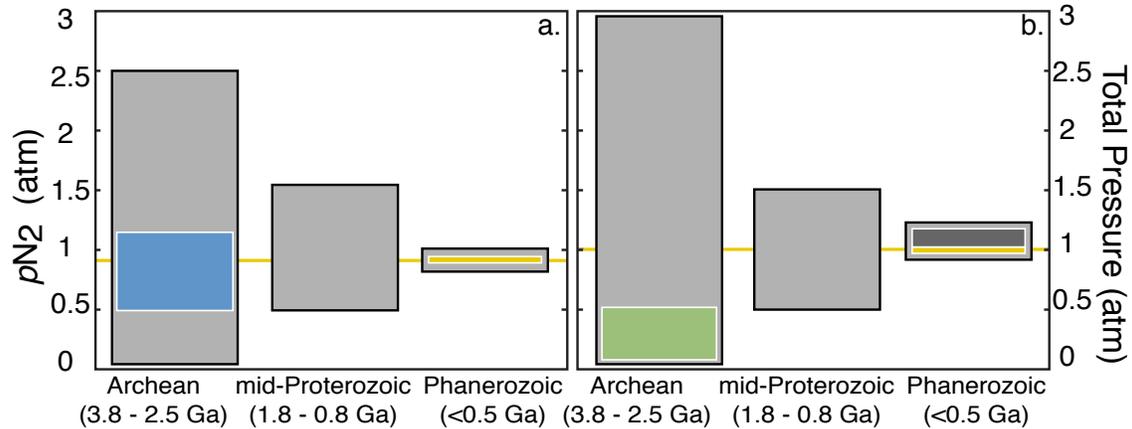

**Figure 4.** *$N_2$ and pressure constraints through Earth history*. For each geological eon, grey boxes represent inclusive ranges for model and proxy based constraints on $pN_2$ (**a**) and total atmospheric pressure (**b**). The colored bars represent preferred ranges corresponding to constraints from specific proxies discussed in the text, including: N isotopes (blue), basalt vesicles (green), and ice core records (light blue). As elsewhere, yellow line denotes modern $pN_2$ and total pressure. Total pressure generally tracks $N_2$ abundance, but the dark grey box in (**b**) represents elevated surface pressure due to very high $O_2$ during the Carboniferous (see **Fig. 1a**). In the Archean, the apparent incompatibility of $pN_2$ and total pressure constraints may be reconciled by considering the temporal separation between the $pN_2$ constraints (3.5-3.0 Ga; Marty et al 2013) and the total pressure constraints (2.7 Ga; Som et al 2016); these complementary datasets may suggest a secular decline in atmospheric pressure during the Archean eon (*e.g.*, Stüeken et al 2016).

What may be a more robust approximation of Archean atmospheric pressure ~2.7 Ga comes from vesicles in basalts (Som et al 2016). The volume of bubbles escaping from lavas is sensitive to the overlying pressure, thus allows determination of atmospheric pressure if elevation can be independently constrained (Sahagian and Maus 1994). In this light, the size distribution of vesicles in Archean basalts erupted at sea level suggest that total pressure was no more than half of modern, implying $pN_2$ was limited to ~60% modern, or ~0.5 atm, after accounting for the absence of $O_2$ in the Archean atmosphere (Som et al 2016).

To date, no other constraints exist for either total atmospheric pressure or $pN_2$ in the Archean. The existing datasets may suggest decreasing atmospheric $pN_2$ and atmospheric pressure throughout the Archean, but additional data will be necessary to improve confidence in this interpretation. Declining $pN_2$ likely requires enhanced N burial in marine sediments in significant excess of modern burial fluxes, either as ammonium ($NH_4^+$) adsorbed to clay minerals or as organic N (Stüeken et al 2016). The former is difficult to reconcile with N isotope evidence for $NH_4^+$ scarcity (Stüeken et al 2015b); the latter is challenged by the C isotope record (*e.g.*, Krissansen-Totton et al 2015). By either scenario, atmospheric $pN_2$ would be expected to recover throughout the Proterozoic because globally extensive denitrification provides an efficient mechanism for returning fixed N to the atmosphere—consistent with existing constraints on the $O_2$ landscape of the Proterozoic ocean (Koehler et al 2017; Olson et al 2016a; Reinhard et al 2016, 2017b; Stüeken et al 2016). That said, $pN_2$ constraints for the Proterozoic are lacking. Despite major uncertainty in the detailed trajectory of $N_2$ evolution, the existing data highlight the likelihood that total atmospheric pressure and $p$N$_2$ have changed through time and the possibility this evolution may not have been unidirectional (Zerkle and Mikhail 2017).

Dinitrogen is also likely to be an important ingredient for habitability elsewhere. As on early Earth, quantifying $p$N$_2$ in an exoplanet's atmosphere will be challenging because N$_2$ lacks standard vibrational-rotational absorption features. However, it is possible that significant levels of N$_2$ ($p$N$_2$ > ~0.5 atm) may be fingerprinted via N$_2$-N$_2$ collision induced absorption (CIA) at 4.3 μm (Schwieterman et al 2015), though planetary characterization at this wavelength may be difficult. Not only would detection of this signal provide context for evaluating an exoplanet's surface pressure and liquid water stability, detection of N$_2$-N$_2$ CIA may preclude abiogenic O$_2$ accumulation on planets with low levels of non-condensing gases (e.g., Wordsworth and Pierrehumbert 2014). In other words, quantifying N$_2$ may be useful for both exoplanet habitability and biosignature studies—strongly motivating continued investigation of the coevolution of life and $p$N$_2$ throughout Earth's history (Stüeken et al 2016).

## 6. Concluding remarks

Not surprisingly, Earth's atmosphere is not static. Nearly every aspect of the atmosphere, both physical and compositional, has changed throughout our 4.5-billion-year history. Some of these changes have been critical to the long-term maintenance of Earth's habitability (*e.g.*, dynamic CO$_2$ adjustment; Walker et al 1981). Others have been a consequence—but not necessarily a direct reflection—of Earth's inhabitation (*e.g.*, protracted oxygenation; Lyons et al 2014). Ultimately, Earth's modern atmosphere is not representative of broad swaths of Earth's history and it is not a terminal state. The early Earth provides many examples of what a habitable planet looks like, and Earth's atmosphere and biosphere will continue to co-evolve with its solid interior and the Sun.

Projecting many million to a few billion years into Earth's future, CO$_2$ will continue to decline in response to continuously increasing solar luminosity (Caldeira and Kasting 1992; Lovelock and Whitfield 1982; but see Lenton and von Bloh 2001). Although major uncertainties exist regarding chemical and climatic regulation mechanisms in Earth history, O$_2$ will likely decline as CO$_2$ and temperature conditions preclude photosynthetic land plants (O'Malley-James et al 2013). The dynamics of a possible return to an anaerobic biosphere are unclear, however, given that continued H escape can permanently oxidize a planet (Catling et al 2001; Zahnle et al 2013). In particular, the potential role of CH$_4$ in regulating climate as CO$_2$ feedbacks eventually fail is unknown. Meanwhile, we might expect that the inevitable reduction in surface pressure that would arise as the result of waning O$_2$ may initially provide a mechanism to extend Earth's habitability by reducing the magnitude of greenhouse warming (Goldblatt et al 2009). If anaerobic biospheres have a greater capacity to draw down N$_2$ (Stüeken et al 2016), however, very low surface pressure may instead accelerate the end of Earth's habitability by exacerbating water loss via H escape in the face of a brightening sun (Wordsworth and Pierrehumbert 2014).

Models for Earth's future atmospheric evolution and the longevity of Earth's biosphere will undoubtedly benefit from continued investigation of the feedbacks that modulated atmospheric composition on the early Earth and proxy reconstructions of atmospheric physicochemical parameters in Earth's past. In particular, improved models for oxygen oases and 'whiffs' of $O_2$ prior to the GOE, as well as $O_2$ stabilization during the mid-Proterozoic, may inform models for the dynamics of eventual planetary deoxygenation and the demise of complexity. Meanwhile, continued examination of $CH_4$ hazes and $CH_4$ cycling in redox-stratified, Proterozoic-like biospheres will be useful for predicting future climate dynamics as $CO_2$ feedbacks collapse. In parallel, studies of the long-term exchange of $N_2$ between Earth's atmosphere and mantle, as well as the role of plate tectonics, will provide critical, yet under appreciated, context for projecting the fate of Earth's biosphere and the limits of habitability on Earth-like planets.

Ultimately, the many Alternative Earths discussed here provide a remarkable catalog of possible planetary states, spanning an extraordinary range of chemical, climatic, and tectonic conditions—and having broad relevance for the diversity of habitable exoplanetary environments. Unraveling the details of Earth system evolution, both past and future, will provide key insight to the mechanics of long-term habitability and will guide our search for life beyond our own planet.

## Acknowledgements

The authors gratefully acknowledge support from the NASA Astrobiology Institute, including support from the Alternative Earths team under Cooperative Agreement Number NNA15BB03A and the Virtual Planetary Laboratory under Cooperative Agreement Number NNA13AA93A. EWS also acknowledges support from the NASA Postdoctoral Program, administered by the Universities Space Research Association.